\pdfoutput=1
\PassOptionsToPackage{hypertexnames=false, unicode, colorlinks=true, linkcolor=linkcolor, citecolor=linkcolor, filecolor=linkcolor, urlcolor=linkcolor, pdfusetitle}{hyperref}
\documentclass[10pt,aps,prd,amsmath,floats,floatfix, twocolumn,
superscriptaddress,nofootinbib,showpacs]{revtex4-1}

\usepackage[T1]{fontenc}
\usepackage[utf8]{inputenc}
\usepackage{lmodern}
\usepackage{orcidlink}
\usepackage{verbatim}
\usepackage[bottom]{footmisc}
\definecolor{linkcolor}{rgb}{0.0,0.3,0.5}
\usepackage[all]{hypcap}
\usepackage{graphicx}
\usepackage{xspace}
\usepackage{amssymb}
\usepackage[normalem]{ulem} 
\usepackage{bm} 
\usepackage{enumitem,amssymb}
\usepackage{lineno}
\usepackage{subfigure}
\usepackage{microtype}
\usepackage[english]{babel}
\usepackage{blindtext}
\usepackage{lineno}
\usepackage{longtable}
\usepackage{xtab}
\usepackage{array}
\usepackage{amssymb}

\begin{document}

\preprint{}

\title{Systematic Uncertainties in Ultralight Boson Constraints from Superradiance}

\author{P.S. Aswathi\orcidlink{0009-0008-1458-3338}}
\author{Ling Sun\orcidlink{0000-0001-7959-892X}}
\affiliation{OzGrav-ANU, Centre for Gravitational Astrophysics, Research School of Physics and Research School of Astronomy \& Astrophysics, The Australian National University, ACT 2601, Australia}


\date{\today}
\begin{abstract}
Ultralight bosons are well-motivated candidates for physics beyond the Standard Model and may constitute part or all of the dark matter. Black-hole superradiance provides a powerful means of probing these particles through their gravitational interaction with rotating black holes. This mechanism has motivated a growing body of constraints from electromagnetic measurements of black-hole spins and, more recently, gravitational-wave observations. Gravitational-wave probes draw on several distinct signatures: the spins of merging binary black holes, gravitational radiation emitted by boson clouds, and potential imprints of cloud-induced changes to binary dynamics.  
The interpretation of these probes depends on method-specific astrophysical and theoretical assumptions, including black hole formation, natal spin, age, accretion history, population priors, and boson self-interactions.
In this paper, we review the principal electromagnetic and gravitational-wave constraints derived from black-hole superradiance, with particular emphasis on their underlying assumptions and systematic uncertainties. 
The distinct observational signatures and systematic uncertainties of these approaches offer complementary perspectives on the viability of ultralight-boson models. In particular, their overlapping coverage around boson masses of $10^{-13}$--$10^{-12}$~eV makes this region particularly informative for comparing constraints, with attention to the assumptions underlying each result.
\end{abstract}


\maketitle
\section{Introduction}
\label{intro}



Ultralight bosons are hypothetical particles motivated by extensions of the Standard Model and can provide viable dark matter candidates~\cite{Jaeckel2010,Essig2013,Hui2017,Bertone2018_1,Bertone2018_2,Oks2021,Baryakhtar2025}. Well-studied examples include axion-like particles arising in string-theoretic compactifications~\cite{Arvanitaki2010,Freitas2021} and massive vector bosons, such as dark photons associated with hidden-sector $U(1)$ gauge symmetries~\cite{Holdom1986,Goodsell2009,Jaeckel2010,Essig2013,Agrawal2020,Fabbrichesi2021}. Their nongravitational couplings to Standard Model particles can be extremely weak, making them challenging to detect through searches that rely on these interactions. Black-hole superradiance provides a complementary probe through gravity alone, extending sensitivity to bosons even when their nongravitational couplings are negligible.

A rotating black hole can superradiantly amplify a mode of a bosonic field of mass $m_b$ when
\begin{equation}    
    0 < \omega_b < m \Omega_H
    \label{eqn:superrad_condition}
\end{equation}
where $\omega_b \approx m_b c^2 / \hbar$ is the angular frequency of the bosonic field, $m$ is the azimuthal quantum number of the bosonic bound state, and $\Omega_H$ is the angular frequency of the black-hole horizon. 
In the linear regime, gravitationally bound modes satisfying this condition grow exponentially. Growth is generally most efficient when the boson's reduced Compton wavelength, $\lambda_C=\hbar/(m_b c)$, is comparable to the black hole's gravitational radius, $r_g=GM/c^2$, where $M$ is the black-hole mass~\cite{Arvanitaki2011,Yoshino2014,Yoshino2015,Arvanitaki2015,Arvanitaki2017,Brito2017,Brito2017_2,Baryakhtar2017,Chan2022,Cardoso2018,Baumann2019,Hannuksela2019,Zhang2019,East2017,East2017_2,East2018,Siemonsen2020,Brito2020,Brito:2020lup}. The growing field extracts energy and angular momentum from the black hole, spinning it down as a macroscopic boson cloud develops~\cite{Zel'Dovich1971,Starobinskii1973,Detweiler1980,Bekenstein1973,Dolan2007,Arvanitaki2011,East2017_2,East2018,Herdeiro2022}. The cloud growth continues until the spin-down brings Eq.~\eqref{eqn:superrad_condition} close to saturation, at which point the instability shuts off. 

Three potentially observable consequences underpin the complementary approaches reviewed in this article. First, superradiance can reduce a black hole's spin toward a critical value set by the boson mass and black-hole mass. Black hole spins above this threshold can constrain the boson mass, provided sufficient time is available for spin-down under the assumed evolution~\cite{Arvanitaki2010,Arvanitaki2011,Brito:2014wla,East2017_2,East2018,Witte:2024drg}. This spin-down signature, visualized as an exclusion region in the black hole mass–spin, or ``Regge,'' plane~\cite{Arvanitaki2015,Brito2017_2}, has long provided a principal probe of ultralight bosons.

Second, boson clouds can emit long-lived, nearly monochromatic gravitational radiation~\cite{East2017_2,East2018,Baryakhtar2017,Isi2019,Siemonsen2020,Baryakhtar2021,Siemonsen2023_2,May2024}, motivating continuous-wave and long-transient searches for individual sources, as well as stochastic-background searches for unresolved populations. Clouds around stellar- and intermediate-mass black holes offer sensitivity to boson masses of roughly $10^{-14}$--$10^{-11}\,\mathrm{eV}$~\cite{Arvanitaki2015,Brito2017_2}, with emission frequencies overlapping the range accessible to current and future ground-based gravitational-wave detectors~\cite{aLIGO,aVirgo,KAGRA,ObservingScenarios,sun2026,acernese2026advancedvirgopluso5,ET:2025xjr,Evans2021}.

Third, if a black hole hosting a boson cloud belongs to a binary, the companion's tidal field can drive resonant transitions between the cloud's energy levels, exchanging energy and angular momentum with the orbit and potentially producing floating or sinking orbits~\cite{Baumann:2019ztm}. As the binary inspirals, the companion can also induce transitions from bound to unbound states, ``ionizing'' the cloud in analogy with the photoelectric effect~\cite{Baumann:2022pkl}. These processes can leave distinctive imprints on the binary's gravitational waveform~\cite{Tomaselli:2024dbw}. 
A complementary waveform analysis constrained scalar environments around observed binaries and reported tentative evidence for such an environment under superradiance-informed priors~\cite{Roy2025}.
Despite these initial studies, this channel remains primarily prospective, with opportunities for next-generation ground-based detectors~\cite{ET:2025xjr,Evans2021} and space-based detectors such as LISA~\cite{DellaMonica:2025zby,Tomaselli:2024dbw}.


These signatures motivate several complementary observational approaches. Electromagnetic constraints use black-hole spins inferred from X-ray continuum-fitting, primarily for stellar-mass systems, or reflection spectroscopy, including the Fe-K$\alpha$ line, for both stellar-mass and supermassive black holes. Sufficiently high inferred spins can constrain boson masses that would produce appreciable spin-down under the assumed evolutionary conditions~\cite{Arvanitaki2015,Baryakhtar2017,Khalaf:2024nwc}. Gravitational-wave observations provide several additional avenues. Merger catalogs support constraints from individual black-hole spin measurements~\cite{aswathi2025,Abac_2025_GW241011_241110,caputo2025} and hierarchical population analyses~\cite{Ng2021,Ning:2026ebu}. Searches also target a stochastic background from unresolved cloud populations~\cite{Tsukada2019,Tsukada2021,Yuan2022,Ellis:2026gwt} and continuous or long-transient emission from individual clouds, through directed searches toward known black holes or selected sky regions and all-sky searches for unknown sources~\cite{O3_all-sky_scalar_bosons,Palomba2019,Dergachev2019,Zhu2020,O3_galactic_center,LIGOScientific:2025csr,Mirasola:2025car}. Resonance and ionization effects on binary inspiral waveforms offer a further, prospective channel~\cite{Baumann:2019ztm,Baumann2019,Baumann2019_2,Baumann:2022pkl}.

\begin{figure*}
    \centering
    \includegraphics[width=\textwidth]{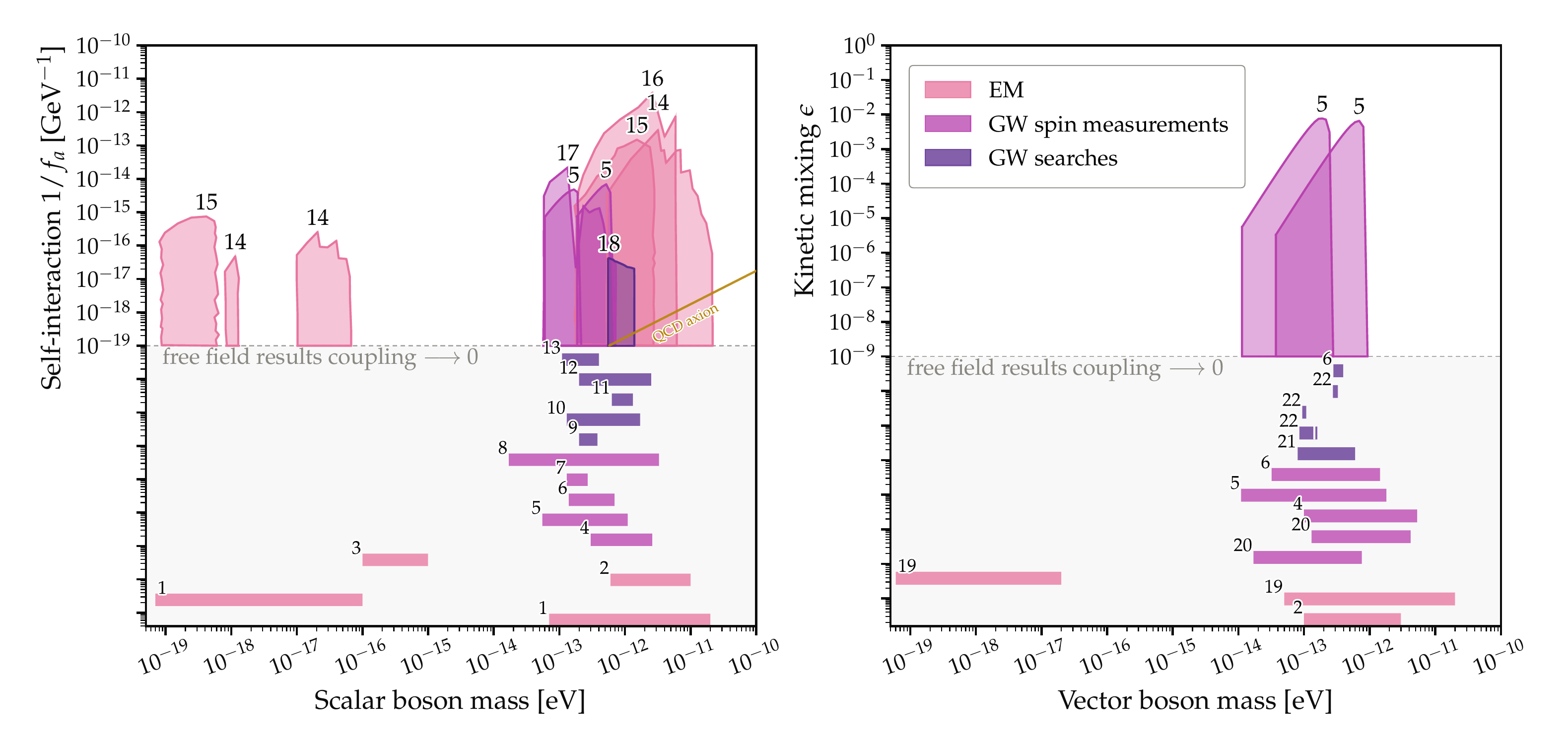}
    \caption{Reported exclusions from black-hole superradiance for scalar (left) and vector (right) ultralight bosons, under the assumptions of the respective analyses. Colors indicate methodology: electromagnetic spin measurements (pink), gravitational-wave spin measurements of individual events or populations (magenta), and searches for gravitational radiation from boson clouds (purple). Filled regions above the dashed line show two-dimensional exclusions in mass and interaction-parameter space. Horizontal bars below the dashed line show mass-only exclusions obtained in the gravity-dominated regime, with no explicit coupling dependence displayed; their vertical positions are chosen solely for legibility and carry no physical meaning. Numbered labels identify the studies listed in the table in Appendix~\ref{app:reference_table}; entries from the same paper share a number. The diagonal line in the scalar panel shows the standard QCD axion mass--decay-constant relation, $m_a f_a \simeq 5.70(6)(4)\times10^{12}\,\mu\mathrm{eV\,GeV}$~\cite{GrillidiCortona:2015jxo}. The exclusions are shown at the confidence or credibility levels reported in the original studies, which vary between results.}
    \label{fig:superradiance_constraints}
\end{figure*}

Figure~\ref{fig:superradiance_constraints} summarizes existing constraints on scalar and vector ultralight bosons, showing the excluded mass windows reported under each analysis's assumptions. These results provide complementary information, but their overlap does not by itself establish a stronger combined constraint. Meaningful comparison requires accounting for differences in astrophysical assumptions, particle models, and statistical treatments, as well as any shared uncertainties. This article makes these dependencies explicit, using the numbered entries in Figure~\ref{fig:superradiance_constraints} to guide the discussion. We review the existing electromagnetic and gravitational-wave constraints, focusing on the systematic uncertainties and modeling choices that determine their scope and interpretation. Our aim is to clarify how the different approaches complement one another, rather than to derive new exclusions.

The remainder of this article is organized as follows. Section~\ref{sec:EM_observations_constraints} reviews electromagnetic constraints and their principal systematics. Section~\ref{sec:GW_spin_constraints} examines gravitational-wave constraints from individual black-hole spin measurements and population analyses, together with their respective systematics. Section~\ref{sec:GW_search} reviews searches for gravitational radiation from boson clouds through stochastic-background, continuous-wave, and long-transient methods. 
Section~\ref{sec:self_interactions} discusses self-interactions and non-gravitational couplings.
Finally, Section~\ref{Sec:conclusions} compares these results in light of their underlying assumptions, assesses their robustness, and discusses their complementarity.


\section{Constraints from electromagnetic observations \label{sec:EM_observations_constraints}}

Most electromagnetic constraints on ultralight bosons rely on superradiant spin-down: sufficiently high measured spins can exclude boson masses that would have reduced the spin within the assumed evolutionary time. Arvanitaki et al.~\cite{Arvanitaki2015} reported scalar exclusions around $6\times10^{-13}$--$2\times10^{-11}\,\mathrm{eV}$ from stellar-mass black-hole spins; Stott and Marsh~\cite{Stott:2018opm} extended spin-based constraints to multifield axion-like-particle (ALP) scenarios; and Hoof et al.~\cite{Hoof:2024quk} applied a Bayesian framework to M33~X-7 and introduced the supermassive black hole IRAS~09149-6206 as a probe. For vectors, Baryakhtar et al.~\cite{Baryakhtar2017} reported exclusions around $5\times10^{-14}$--$2\times10^{-11}\,\mathrm{eV}$ from stellar-mass black holes and lower-confidence constraints over $6\times10^{-20}$--$2\times10^{-17}\,\mathrm{eV}$ from supermassive systems. Using the observed stability of stellar-mass black-hole accretion disks, Cardoso et al.~\cite{Cardoso2018} derived dark-photon limits around $10^{-13}$--$3\times10^{-12}\,\mathrm{eV}$ and, including higher azimuthal modes, ALP limits around $6\times10^{-13}$--$10^{-11}\,\mathrm{eV}$. They also explored supermassive-black-hole mass and spin distributions as probes of lighter bosons. As a complementary approach, Khalaf et al.~\cite{Khalaf:2024nwc} proposed comparing continuum-fitting and reflection-based spin estimates for the same source to probe the gravitational contribution of a boson cloud.

These constraints all rest on measured BH spin, obtained via two X-ray spectroscopic techniques: continuum-fitting, which models the thermal disk spectrum~\cite{McClintock:2013vwa}, and reflection spectroscopy, which models relativistically broadened reflection features, notably the Fe-K$\alpha$ line~\cite{Reynolds:2013qqa,Reynolds:2013rva,Draghis:2022ngm}. 
Both infer the same underlying quantity, the disk's inner radius, assumed to coincide with the innermost stable circular orbit (ISCO) $R_{\mathrm{ISCO}}$, which decreases monotonically with BH spin. They use different signatures: continuum-fitting uses the disk temperature and brightness, requiring independent knowledge of distance, inclination, and mass to infer the radius and spin; reflection spectroscopy uses the Doppler- and gravitationally-redshifted shape of the iron line, with its extension toward lower energies providing sensitivity to emission from the innermost disk.

Both methods share a potential systematic: standard implementations identify the ISCO with the inner edge of the emission, whereas simulations find non-zero disk density and temperature inside it. Neglecting this residual ``plunging-region'' emission can bias both types of spin estimate, and its inclusion has been proposed to explain excess emission in spectra of sources such as MAXI~J1820+070 and MAXI~J0637-430~\cite{Mummery:2023tgh,Dong:2023bbd,Wilkins:2020pgu,Noble:2010mm,Zhu:2012vf,Lancova:2022bha,Fabian:2020lvv,Lazar:2021ccv,Mummery:2024mrq,Mummery:2024tyo,Witte:2024drg}.

Continuum-fitting's error budget can be dominated by disk-atmosphere modelling rather than by its inner boundary condition: incomplete photon thermalization can mimic spin-driven spectral hardening, and adopting colour-correction uncertainties ($\pm0.2$--$0.3$) can set the error budget for most measurements examined by Salvesen and Miller~\cite{Salvesen:2020bds}. It also commonly assumes spin-orbit alignment and, because it requires directly observable thermal disk emission, is primarily applied to stellar-mass BHs~\cite{McClintock:2013vwa,Arvanitaki2015}. One exception is the tidal disruption event 3XMM~J215022.4$-$055108, for which a slim-disk fit ($a=0.80^{+0.12}_{-0.02}$) yielded the first superradiance constraint at intermediate BH masses, around $10^{-16}$--$10^{-15}\,\mathrm{eV}$, though from a single source and under assumptions about its host and evolutionary timescale~\cite{Wen:2021yhz}. Reflection spectroscopy, by contrast, probes the dimensionless ratio $R_{\mathrm{ISCO}}/r_g$ without requiring independent mass and distance measurements. It therefore applies across the BH mass range, including supermassive BHs~\cite{Arvanitaki2015}, but is subject to its own degeneracies involving iron abundance, line-of-sight absorption, and soft-excess modelling, as well as instrumental biases such as photon pile-up in bright sources. These effects can shift the inferred spin~\cite{Reynolds:2013qqa}.

Comparisons between the methods can reveal potential systematics. Among BH X-ray binaries with both reflection- and continuum-fitting spin measurements, most agreed in the sample reviewed in Ref.~\cite{Reynolds:2013qqa}, but two were discrepant; in at least one such case, the lower estimate was adopted conservatively when setting a bound~\cite{Arvanitaki2015}. This cross-check was flagged in early reviews~\cite{McClintock:2013vwa,Reynolds:2013qqa} and has more recently been proposed as a dedicated observational test for the presence of a boson cloud~\cite{Khalaf:2024nwc}.

A further caveat concerns any single high-spin measurement. Post-formation accretion can spin a BH up, so early superradiant spin-down need not leave a signature today. For Cygnus~X-1, M33~X-7, and LMC~X-1, the Eddington-limited accretion timescale needed to reach the observed spin exceeds the estimated system age by factors of several, favoring a natal origin~\cite{McClintock:2013vwa}. More recent evolutionary modelling has explored spin-up through highly conservative, super-Eddington accretion, although reproducing all observed properties of Cygnus~X-1 remains challenging~\cite{Xing2025}. The spin measurement itself can also be model-sensitive: for Cygnus~X-1, alternative spectral models including a ``warm'' corona layer allow low spins consistent with zero, compared with earlier estimates of $a\gtrsim0.9985$, although the applicability of this alternative model remains uncertain~\cite{Zdziarski_2024a}. Excluding a boson mass from a single source therefore requires comparing the superradiant spin-down timescale with the source age and relevant spin-up timescales. Arvanitaki et al.~\cite{Arvanitaki2015} also account for tidal perturbations from a companion star and delays in further spin-down when boson self-interactions temporarily suppress the growth of additional cloud modes. In that analysis, stellar-mass exclusions were considered relatively robust, while supermassive-BH bounds were treated as preliminary, relying on a single spin-measurement method, uncertain ages, and uncertain estimates of how often infalling compact objects could disturb the cloud~\cite{Arvanitaki2015}.

Finally, most of the bounds reported above assume a free, non-self-interacting field, whereas most physically motivated candidates carry a self-interaction set by an independent parameter; we defer this discussion, and its impact on the exclusions above and in the sections that follow, to Section~\ref{sec:self_interactions}.

\section{Constraints from black-hole spins inferred from gravitational-wave observations\label{sec:GW_spin_constraints}}


Gravitational-wave observations of compact binary mergers offer a route to the same Regge-plane exclusion logic without relying on X-ray accretion models. The observable is the black hole's spin, inferred from the merger waveform. The most direct application uses the spin posterior of a single, well-measured event: a spin incompatible with efficient superradiant spin-down, given an assumed minimum age, can exclude the corresponding boson mass, as in the electromagnetic case but without dependence on a disk or corona model.
Applied to GW190517\_055101 and GW190426\_152155, this approach excluded dark photon masses of $[0.17,7.6]\times10^{-13}$~eV and $[1.3,42]\times10^{-13}$~eV, respectively, assuming a formation-to-merger timescale of $10^7$ years~\cite{Ghosh:2021zuf}. More recently, an analysis of GW231123\_135430 and GW190517\_055101 adopted a shorter minimum BH age of $10^5$ years, excluding scalar masses of $[0.55,3.3]\times10^{-13}$~eV and $[2.2,11]\times10^{-13}$~eV, respectively, and vector masses of $[0.11,4.9]\times10^{-13}$~eV and $[0.43,18]\times10^{-13}$~eV at 90\% confidence~\cite{aswathi2025}. The different windows obtained for GW190517\_055101 illustrate the combined influence of age assumptions, mode inclusion, and statistical treatment. An independent analysis of GW231123\_135430 considering different formation channels found a comparable excluded axion-mass range of approximately $[0.6,5]\times10^{-13}$~eV, with the exclusion depending also on the self-interaction strength~\cite{caputo2025}.
The LIGO--Virgo--KAGRA Collaboration's analysis of GW241011\_233834 and GW241110\_124123 used the well-measured primary spin of GW241011\_233834 to exclude scalar and vector boson masses of $[0.3,2.6]\times10^{-12}$~eV and $[0.1,5.3]\times10^{-12}$~eV, respectively, at 90\% credibility, assuming a BH age of $10^5$ years~\cite{Abac_2025_GW241011_241110}. Three additional high-spin events in the second part of the fourth observing run (O4b), GW240515\_005301, GW241113\_163507, and GW241225\_082815, yielded exclusions spanning $[1.39,6.94]\times10^{-13}$~eV for scalars and $[0.32,14.4]\times10^{-13}$~eV for vectors at 90\% confidence under the same age assumption. These intervals represent the union of the three individual-event exclusions~\cite{LIGOScientific:2026pwx}.

This individual-event approach inherits waveform-model uncertainties through its dependence on the measured spin posterior. Different signal models can disagree non-negligibly on inferred component spins, particularly for the secondary in asymmetric-mass systems, though the lowest-mass end of the exclusions studied in Ref.~\cite{aswathi2025} is relatively robust to this systematic. A related caveat is that the waveform model, NRSur7dq4, though trained only on numerical-relativity simulations with component spins $\chi_{1,2}\leq0.8$~\cite{Varma2019}, has been used to infer larger spins in several parameter-estimation analyses~\cite{GWTC-4_results,GWTC-5}. For events with posterior support above this threshold, particularly GW241225\_082815, this extrapolation is a potential source of waveform-model uncertainty in the inferred spins and resulting superradiance constraints.
GW231123\_135430 illustrates additional event-specific uncertainties. Its inferred properties show waveform-model dependence~\cite{GW231123}, and possible biases from non-Gaussian noise have also been investigated~\cite{Ray:2025rtt}. Its primary mass lies within or above the predicted pair-instability mass gap, leaving its formation channel uncertain and complicating the astrophysical justification for an assumed minimum age~\cite{GW231123}. Nevertheless, the excluded boson-mass range varies only modestly over the $10^5$--$10^7$ years explored in Ref.~\cite{aswathi2025}, provided the inferred masses and spins are reliable.

A complementary strategy combines spin measurements across a binary black hole catalog in a hierarchical Bayesian framework~\cite{Ng:2019jsx}. Using simulated catalogs, Ref.~\cite{Ng:2019jsx} showed that $\mathcal{O}(100)$ high signal-to-noise detections could exclude (or confirm) non-interacting scalar bosons with masses of roughly $10^{-13}$--$10^{-12}$~eV.  Rather than testing individual spins against a critical value, this approach compares the observed mass and spin population with models incorporating superradiant spin-down, under assumptions about natal spins, masses, and evolutionary timescales. In Ref.~\cite{Fernandez:2019qbj}, a hierarchical analysis was applied to the ten binary black hole mergers then available from the first two LIGO--Virgo observing runs (O1 and O2). Despite a mild, model-dependent preference for superradiance, the evidence was insufficient for a statistically significant conclusion. Applied to GWTC-2, a population analysis disfavored scalar boson masses in the range $[1.3,2.7]\times10^{-13}$~eV, assuming a formation-to-merger timescale of $10^7$ years, with the exclusion driven mainly by GW190517\_055101 and GW190412~\cite{Ng2021}. 
A related analysis explored the dependence on natal-spin and formation-to-merger timescale assumptions, finding only weak, model-dependent preferences for particular scalar masses~\cite{Cheng2023}.
A more recent analysis using the substantially larger GWTC-5 catalog reported a broader exclusion of scalar masses $[0.17,33]\times10^{-13}$~eV at 95\% credibility under its population and evolutionary assumptions~\cite{Ning:2026ebu}.
Allowing phenomenological spin variations, an analysis of GWTC-3.0 through GWTC-5.0 favored scalar masses near $7\times10^{-13}\,\mathrm{eV}$~\cite{Kou2026}, within regions excluded by Refs.~\cite{Ning:2026ebu,Abac_2025_GW241011_241110}. Whether environmental effects can reconcile this preference with GW241011\_233834's high spin remains to be further investigated.

The population-level approach combines information from many events but introduces additional assumptions about the underlying black hole natal spin and mass distributions. A further consideration is that superradiance can itself alter the merger population. Ref.~\cite{Payne:2021ahy} showed that, for boson masses $2\times10^{-14}$--$2\times10^{-13}$~eV, superradiant spin-down can reduce merger kicks and increase the retention of merger remnants in nuclear star clusters, allowing more hierarchical mergers. Population-level sensitivity also depends on catalog size and composition. The broader exclusion reported using GWTC-5 highlights the information available from larger catalogs, although differences in modelling and statistical assumptions also enter the comparison with GWTC-2~\cite{Ng2021,Ning:2026ebu}.

Both approaches, however, share two further modelling requirements. Each must specify the time available for superradiant spin-down, through an assumed age or age distribution. This timescale is not directly measured for merging binaries, and longer assumed times generally strengthen the exclusions~\cite{aswathi2025,Ng2021,Ning:2026ebu}. Both must also account for whether accretion or environmental effects could compete with or interrupt superradiant spin-down. The relevant timescale may therefore be the time since the last significant spin-up rather than since formation~\cite{aswathi2025}. Unmodelled effects can alter an individual-event exclusion and, in a population analysis, may also bias the inferred natal spin distribution and resulting boson constraints. Electromagnetic sources can provide additional information about age and accretion history, as discussed in Section~\ref{sec:EM_observations_constraints}.

\section{Gravitational-wave searches for boson clouds}
\label{sec:GW_search}

A different observational strategy searches for gravitational radiation from the boson clouds themselves, rather than for their imprint on a black hole's spin. The superposition of many unresolved, long-lived, quasi-monochromatic signals from these clouds can form a stochastic gravitational-wave background. Under optimistic assumptions about the BH population, searches for such background signals have excluded scalar boson masses in the range $[2.0,3.8]\times10^{-13}$~eV~\cite{Tsukada2019} and, subsequently, vector boson masses in the range $[0.8,6.0]\times10^{-13}$~eV~\cite{Tsukada2021}, both at 95\% credibility. A later analysis using data from the first three observing runs (O1--O3) and including higher cloud modes further constrained the scalar case~\cite{Yuan2022}.

Because the predicted stochastic background is sourced by an assumed population of black holes, these constraints depend on the formation rate, mass function, redshift distribution, and natal spin distribution, together with the timescales for cloud growth and emission~\cite{Brito2017,Brito2017_2}. These assumptions overlap with those entering hierarchical spin analyses, but affect the predicted observables differently. In particular, a population with fewer rapidly spinning black holes generally produces a weaker background, so exclusions can shrink substantially or disappear under predominantly low-spin assumptions~\cite{Tsukada2019}. Natal-spin uncertainty therefore affects both stochastic-background and population-level spin constraints, although through different mechanisms.

Continuous-wave searches target the nearly monochromatic signal from an individual cloud directly. Blind all-sky searches scan the sky without assuming a source direction. Searches using earlier LIGO data placed strain upper limits that informed constraints on scalar boson clouds~\cite{Palomba2019,Dergachev2019}. A subsequent all-sky search using the full O3 observing run set exclusion regions in the boson-mass--black-hole-mass plane for specified source distances, initial spins, and cloud ages~\cite{O3_all-sky_scalar_bosons}. These searches report strain upper limits as a function of frequency, which can be interpreted under assumed source properties or population models. For example, Zhu et al.~\cite{Zhu2020} simulated a population of isolated Galactic black holes hosting clouds and assessed the detectability of the resulting signals in earlier all-sky searches. Under their population assumptions, the non-detections disfavored scalar masses of approximately $2\times10^{-13}$--$2.5\times10^{-12}$~eV, with additional uncertainty where signals overlap densely in frequency.

A second category targets a specific sky direction expected to host a dense population of potential sources, most notably the Galactic Center, without identifying any individual source within it. Such directed searches can gain sensitivity because no sky-position scan is required. An O3 search of this kind interpreted its strain upper limits as constraints on boson clouds at the Galactic Center, assuming a source distance, BH spin, and cloud age rather than using measured properties of a single known source~\cite{O3_galactic_center}.

Source-directed searches target electromagnetically identified black holes with independently constrained masses, distances, and, in some cases, ages, allowing exclusions based on an individually characterized source. A directed search of Cygnus~X-1 disfavored scalar boson masses of $[6.3,13.2]\times10^{-13}$~eV, assuming negligible self-interactions, an initial BH spin of $0.99$, and an age of $10^5$ years~\cite{O2_CygnusX1_scalar_bosons}; its self-interacting extension and subsequent reanalysis are discussed in Section~\ref{sec:self_interactions}. A more recent search using data from the first part of the fourth observing run excluded vector boson masses spanning $[0.85,1.59]\times10^{-13}$~eV at 95\% confidence, apart from a narrow gap at $[1.44,1.45]\times10^{-13}$~eV, assuming an initial BH spin above $0.5$ and an estimated age of $(6.2\pm1.8)\times10^6$ years~\cite{LIGOScientific:2025csr}.

A distinct class of directed search targets the merger remnants of individual gravitational-wave events, using a hidden-Markov-model tracking method~\cite{Jones2023,Jones2025}. Here the remnant's formation time is fixed by the observed merger, removing the age uncertainty that affects Galactic searches. The trade-off is the much greater distance of these remnants, which limits current sensitivity. At these distances, the long-duration signals from scalar clouds are generally expected to be too weak for current detectors~\cite{Isi2019,Ghosh:2018gaw}. Vector clouds can radiate more strongly over shorter timescales, offering better prospects for detection with existing instruments~\cite{Jones2023}. Targeting the remnants of GW231123\_135430 and GW230814\_230901, this method marginally disfavored vector boson masses of $[0.94,1.08]\times10^{-13}$~eV and $[2.75,3.28]\times10^{-13}$~eV, respectively, at 30\% confidence~\cite{LIGOScientific:2025csr}. A more recent search targeting the remnants of GW250114\_082203 and GW250207\_115645 found no evidence for a signal and disfavored vector boson masses spanning $[2.80,3.95]\times10^{-13}$~eV at greater than 90\% confidence (the union of the two individual-search exclusions)~\cite{LIGOScientific:2026pwx}.

Continuous-wave searches carry systematics distinct from the other gravitational-wave-based methods discussed above. Beyond the source and population assumptions already noted, many reported exclusions assume negligible boson self-interactions. Section~\ref{sec:self_interactions} discusses how self-interactions and other non-gravitational couplings can alter these exclusions. Instrumental spectral disturbances also complicate candidate follow-up. In the O3 scalar all-sky search, the narrow frequency-derivative range made instrumental artifacts harder to identify than in standard continuous-wave searches covering broader spin-down or spin-up ranges~\cite{O3_all-sky_scalar_bosons}. Sensitivity also depends on whether the search accommodates the signal's frequency evolution, including any changes induced by self-interactions. A mismatch can reduce sensitivity, although the self-interacting signals examined in Ref.~\cite{Collaviti2024} remained within the frequency-drift range covered by the earlier Cygnus~X-1 search.


\section{Self-interactions and non-gravitational couplings}
\label{sec:self_interactions}

Many exclusions reviewed in Sections~~\ref{sec:EM_observations_constraints}--\ref{sec:GW_search} assume a free boson field interacting only through gravity. Physically motivated candidates can depart from this limit in two distinct ways: self-interactions, generically present for axion-like fields and controlled by parameters such as the decay constant $f_a$, and non-gravitational couplings to Standard Model fields, such as the kinetic mixing $\epsilon$ between a dark photon and the ordinary photon. These interactions can modify cloud growth, depletion, and radiation, weakening or removing some free-field exclusions while also introducing additional observational signatures. This section reviews how the constraints discussed above have been re-examined when these effects are included, and where further study is needed.

Sufficiently strong self-interactions can alter the superradiant growth assumed in the free-field treatment, with dynamics that depend on the field and interaction model. For scalar fields, early work proposed a ``bosenova'', a rapid, partial collapse and ejection of the cloud driven by attractive self-interactions~\cite{Arvanitaki2011,Yoshino:2012kn}. Later studies incorporating energy exchange between bound states, absorption by the BH, and emission to infinity find that the cloud can instead reach a quasi-equilibrium below the bosenova threshold over much of the parameter space studied~\cite{Baryakhtar2021,Collaviti2024}. 
A relativistic treatment including multiple modes also finds that mode coupling can saturate cloud growth below the bosenova threshold in the regimes studied~\cite{Omiya2023MultipleModes}.
For vector fields, a bosenova can occur in the Higgs-Abelian model, where vortex-string formation and an explosive release of energy temporarily disrupt the cloud, after which superradiant growth can resume~\cite{East2022Vortex,EastHuang2022Vortex}.
Reduced cloud occupation can suppress gravitational-wave emission and slow BH spin-down, weakening exclusions derived in the free-field limit. Such observations therefore constrain combinations of boson mass and self-interaction strength, rather than the mass alone. 

The extent to which self-interactions have been quantified varies across the constraints reviewed above. For electromagnetic spin constraints (Section~\ref{sec:EM_observations_constraints}), Witte \& Mummery~\cite{Witte:2024drg} re-derived stellar-mass BH exclusions, including Cygnus~X-1, using a Bayesian framework and coupled evolution of states with principal quantum number $n\leq5$. For Cygnus~X-1, their limit-setting procedure extended sensitivity to $f_a$ values roughly an order of magnitude smaller than the earlier timescale-based prescription of Ref.~\cite{Baryakhtar2021}. Their comparisons also showed that treatments imposing only a bosenova threshold can extend exclusions to $f_a$ values up to an order of magnitude smaller than treatments including level mixing. Hoof et al.~\cite{Hoof:2024quk}, already introduced in Section~\ref{sec:EM_observations_constraints} for their M33~X-7 and IRAS~09149-6206 analysis, developed a complementary Bayesian framework using BH mass and spin posterior samples to constrain boson masses and self-interactions jointly, while comparing alternative prescriptions for cloud evolution.

For individual-event gravitational-wave spin constraints (Section~\ref{sec:GW_spin_constraints}), self-interactions and other couplings have also been considered. The GW231123\_135430/GW190517\_055101 analysis~\cite{aswathi2025} identifies the interaction strengths for which its free-field exclusions remain applicable to axion and dark-photon models. An independent GW231123\_135430 analysis~\cite{caputo2025} derives exclusions in the mass--decay-constant plane, reaching $f_\Phi\sim10^{14}$~GeV under its formation-channel assumptions. Ref.~\cite{Xie:2025npy} separately examines GW190412 and GW190517\_055101 within its single-mode treatment, illustrating how self-interactions can substantially weaken spin-based constraints.

For population-level, hierarchical Bayesian spin constraints (Section~\ref{sec:GW_spin_constraints}), the GWTC-2~\cite{Ng2021} and GWTC-5~\cite{Ning:2026ebu} analyses model superradiant spin-down in the free-field limit, without explicitly including self-interactions in the population evolution. A similar assumption underlies many of the mass exclusions derived from all-sky, region-directed, and stochastic-background searches in Section~\ref{sec:GW_search}. Extending these interpretations to interacting models would help establish how their sensitivity depends on the additional coupling parameters.
 
For the direct gravitational-wave search channels (Section~\ref{sec:GW_search}), the Cygnus~X-1 directed continuous-wave search reported a self-interacting exclusion alongside its free-field result. In addition to the free-field range $[6.3,13.2]\times10^{-13}$~eV, it inferred an excluded window at $[9.6,15.5]\times10^{-13}$~eV for $f_a\sim10^{15}$~GeV using a bosenova-based model in a string-axiverse scenario~\cite{O2_CygnusX1_scalar_bosons}. Collaviti et al.~\cite{Collaviti2024} subsequently reinterpreted the same search limits using cloud dynamics that include energy exchange between bound states. For an assumed age of $10^5$~yr, they obtained a broadly consistent exclusion spanning approximately $[6,14]\times10^{-13}$~eV where self-interactions are negligible, but found the predicted emission at $f_a\sim10^{15}$~GeV too weak to support the earlier exclusion. The associated frequency drift remained within the original search's capabilities.

A distinct caveat concerns bosons coupled to Standard Model fields, most concretely through kinetic mixing $\epsilon$ between a dark photon and the ordinary photon. This coupling introduces an additional electromagnetic energy-loss channel. For sufficiently small mixing, cloud growth and subsequent depletion remain close to the purely gravitational evolution, so the corresponding constraints remain applicable under their original assumptions. At larger mixing, the cloud can source a rotating electromagnetic field that drives a pair-production cascade, forming a magnetized plasma in which turbulent magnetic reconnection powers electromagnetic emission~\cite{Siemonsen2023_2}. If these losses compete with gravitational-wave emission, they can shorten the cloud's lifetime and alter its gravitational-wave signal, requiring the search sensitivity to be reassessed. Superradiant growth can remain approximately unchanged when electromagnetic losses are slow compared with growth, but this cannot be assumed throughout the interacting parameter space.

This regime has already motivated a dedicated search~\cite{Mirasola:2025car}, using pulsar-catalog sources as candidates for old Galactic black holes hosting electromagnetically luminous dark-photon clouds. Such systems could resemble millisecond radio pulsars while also emitting continuous gravitational waves. No signal was found in O3 data. Interpreted under assumptions about the Galactic BH population and the clouds' radio emission, the resulting limits disfavor a discrete set of dark-photon masses between approximately $10^{-13}$ and $10^{-12}$~eV for kinetic mixing $\epsilon\sim10^{-9}$--$10^{-7}$.

Self-interactions and non-gravitational couplings can weaken existing exclusions while also introducing additional observational signatures and search opportunities. The Cygnus~X-1 case illustrates how an exclusion can change when the assumed cloud dynamics are revised. Free-field bounds therefore apply where additional interactions do not significantly alter the relevant evolution or signal. Dedicated treatments now exist for electromagnetic spin constraints, individual-event spin constraints from gravitational-wave measurements, and source-directed continuous-wave searches, with the scope of these treatments varying across models and observational channels.


\section{Conclusions}
\label{Sec:conclusions}

In this paper, we have reviewed constraints on ultralight bosons from black-hole superradiance using electromagnetic spin measurements, gravitational-wave observations of individual merging black holes and their populations, and direct searches for radiation from boson clouds. These approaches probe overlapping regions of parameter space through distinct observational signatures. Electromagnetic spin measurements depend on accretion-disk and X-ray spectral modelling, while gravitational-wave spin measurements depend on waveform modelling. Both require assumptions about the time available for superradiance and competing spin-evolution processes, with population analyses additionally depending on the underlying mass and natal-spin distributions. Direct cloud searches depend on cloud evolution, source properties or population assumptions, and the signal model. These methods therefore offer complementary perspectives, with both distinct and shared systematic uncertainties.

The overlapping coverage, particularly around $10^{-13}$--$10^{-12}$~eV, makes this region informative for comparing results across observational channels. Such comparisons can identify where exclusions persist across different assumptions and where their interpretation remains sensitive to particular modelling choices. Assessing their combined significance requires accounting for these assumptions and any shared uncertainties. Differences between reported exclusions can motivate targeted tests of the astrophysical and particle models used to derive them.

Looking ahead, growing gravitational-wave catalogs, improved electromagnetic spin measurements and spectral models, and increasing detector sensitivity will expand the reach of these approaches. Further work on black-hole formation and evolution, boson self-interactions, and non-gravitational couplings will clarify how the resulting constraints apply to specific models. Considered together with their assumptions made explicit, these complementary probes provide a broader view of ultralight-boson parameter space and guide future searches across electromagnetic and gravitational-wave observations.
 

\acknowledgments{}
The authors thank Will East for the helpful discussion and internal review.
This material is based upon work supported by NSF's LIGO Laboratory which is a major facility fully funded by the National Science Foundation.
This work is supported by the Australian Research Council (ARC) Centre of Excellence for Gravitational Wave Discovery (OzGrav), Project Number CE230100016. L.S. is also supported by the ARC Discovery Early Career Researcher Award, Project Number DE240100206. 

\bibliography{references.bib}{}
\clearpage
\appendix
\onecolumngrid
\section{Reference table for Figure~\ref{fig:superradiance_constraints}}
\label{app:reference_table}

\begin{center}
\footnotesize
\setlength{\tabcolsep}{4pt}
\begin{longtable}{|l|l|l|l|l|}
\hline
 & Reference & System(s) & Boson & Excluded mass range \\
\hline
\endfirsthead
\hline
 & Reference & System(s) & Boson & Excluded mass range \\
\hline
\endhead
\hline
\endfoot
\hline
\endlastfoot

\parbox[t]{0.6cm}{\centering 1} & \parbox[t]{4.3cm}{Stott \& Marsh (2018) \cite{Stott:2018opm}} & \parbox[t]{4.0cm}{Stellar BHs \newline SMBHs} & \parbox[t]{1.9cm}{\centering Scalar} & \parbox[t]{5.4cm}{$7\times10^{-14}$--$2\times10^{-11}$~eV (stellar) \newline $7\times10^{-20}$--$1\times10^{-16}$~eV (SMBH)} \\
\hline

\parbox[t]{0.6cm}{\centering 2} & \parbox[t]{4.3cm}{Cardoso, Dias, Hartnett, Middleton, Pani \& Santos (2018) \cite{Cardoso2018}} & \parbox[t]{4.0cm}{Cyg X-1 (14~yr), LMC X-3 (26~yr) \newline disk-stability method} & \parbox[t]{1.9cm}{\centering Scalar, Vector} & \parbox[t]{5.4cm}{Scalar: $6\times10^{-13}$--$1\times10^{-11}$~eV \newline Vector: $1\times10^{-13}$--$3\times10^{-12}$~eV} \\
\hline

\parbox[t]{0.6cm}{\centering 3} & \parbox[t]{4.3cm}{Wen, Jonker, Stone \& Zabludoff (2021) \cite{Wen:2021yhz}} & \parbox[t]{4.0cm}{IMBH J2150 (TDE-disk spin measurement)} & \parbox[t]{1.9cm}{\centering Scalar} & \parbox[t]{5.4cm}{$1\times10^{-16}$--$1\times10^{-15}$~eV} \\
\hline

\parbox[t]{0.6cm}{\centering 4} & \parbox[t]{4.3cm}{Abac et al. (2025) \cite{Abac_2025_GW241011_241110}} & \parbox[t]{4.0cm}{GW241011\_233834 primary BH spin} & \parbox[t]{1.9cm}{\centering Scalar, Vector} & \parbox[t]{5.4cm}{Scalar: $3\times10^{-13}$--$2.6\times10^{-12}$~eV \newline Vector: $1\times10^{-13}$--$5.3\times10^{-12}$~eV \newline (t$_{\rm age}=10^5$~yr)} \\
\hline

\parbox[t]{0.6cm}{\centering 5} & \parbox[t]{4.3cm}{Aswathi, East, Siemonsen, Sun \& Jones (2025) \cite{aswathi2025}} & \parbox[t]{4.0cm}{GW231123\_135430, GW190517\_055101 primary BH spin (union of individual-event exclusions) \newline GW231123\_135430, GW190517\_055101 primary BH spin (2-D per-event)} & \parbox[t]{1.9cm}{\centering Scalar, Vector} & \parbox[t]{5.4cm}{1-D: Scalar $5.5\times10^{-14}$--$1.1\times10^{-12}$~eV \newline 1-D: Vector $1.1\times10^{-14}$--$1.8\times10^{-12}$~eV \newline 2-D: GW231123\_135430 scalar $6.0\times10^{-14}$--$2.2\times10^{-13}$~eV \newline 2-D: GW190517\_055101 scalar $1.9\times10^{-13}$--$6.9\times10^{-13}$~eV \newline 2-D: GW231123\_135430 vector $1.1\times10^{-14}$--$2.8\times10^{-13}$~eV \newline 2-D: GW190517\_055101 vector $3.7\times10^{-14}$--$9.3\times10^{-13}$~eV \newline (t$_{\rm age}=10^5$~yr)} \\
\hline

\parbox[t]{0.6cm}{\centering 6} & \parbox[t]{4.3cm}{Abac et al. (2026) \cite{LIGOScientific:2026pwx}} & \parbox[t]{4.0cm}{GW240515\_005301, GW241113\_163507, GW241225\_082815 primary BH spin (union of individual-event exclusions) \newline GW250114\_082203, GW250207\_115645 remnants (continuous-wave)} & \parbox[t]{1.9cm}{\centering Scalar, Vector} & \parbox[t]{5.4cm}{1-D: Scalar $1.39\times10^{-13}$--$6.94\times10^{-13}$~eV \newline 1-D: Vector $0.32\times10^{-13}$--$14.4\times10^{-13}$~eV \newline (t$_{\rm age}=10^5$~yr) \newline CW (vector): $2.80\times10^{-13}$--$3.95\times10^{-13}$~eV} \\
\hline


\parbox[t]{0.6cm}{\centering 7} & \parbox[t]{4.3cm}{Ng, Vitale, Hannuksela \& Li (2021) \cite{Ng2021}} & \parbox[t]{4.0cm}{GWTC-1/GWTC-2 population (45 BBHs)} & \parbox[t]{1.9cm}{\centering Scalar} & \parbox[t]{5.4cm}{$1.3\times10^{-13}$--$2.7\times10^{-13}$~eV} \\
\hline

\parbox[t]{0.6cm}{\centering 8} & \parbox[t]{4.3cm}{Ning, Safdi \& Welch (2026) \cite{Ning:2026ebu}} & \parbox[t]{4.0cm}{GWTC-5 population ($N=257$ mergers)} & \parbox[t]{1.9cm}{\centering Scalar} & \parbox[t]{5.4cm}{$1.7\times10^{-14}$--$3.3\times10^{-12}$~eV} \\
\hline

\parbox[t]{0.6cm}{\centering 9} & \parbox[t]{4.3cm}{Tsukada, Callister, Matas \& Meyers (2019) \cite{Tsukada2019}} & \parbox[t]{4.0cm}{Cosmological stochastic GW background \newline (Advanced LIGO O1)} & \parbox[t]{1.9cm}{\centering Scalar} & \parbox[t]{5.4cm}{$2.0\times10^{-13}$--$3.8\times10^{-13}$~eV} \\
\hline

\parbox[t]{0.6cm}{\centering 10} & \parbox[t]{4.3cm}{Yuan, Jiang \& Huang (2022) \cite{Yuan2022}} & \parbox[t]{4.0cm}{Cosmological stochastic GW background \newline (Advanced LIGO/Virgo O1--O3)} & \parbox[t]{1.9cm}{\centering Scalar} & \parbox[t]{5.4cm}{$1.3\times10^{-13}$--$1.7\times10^{-12}$~eV \newline (95\% C.L.; optimistic natal spins uniform over $[0.5,1]$; other spin distributions give different intervals)} \\
\hline

\parbox[t]{0.6cm}{\centering 11} & \parbox[t]{4.3cm}{Sun, Brito \& Isi (2020) \cite{O2_CygnusX1_scalar_bosons}} & \parbox[t]{4.0cm}{Cygnus X-1 (directed continuous-wave search)} & \parbox[t]{1.9cm}{\centering Scalar} & \parbox[t]{5.4cm}{$6.3\times10^{-13}$--$13.2\times10^{-13}$~eV \newline (t$_{\rm age}=10^5$~yr)} \\
\hline

\parbox[t]{0.6cm}{\centering 12} & \parbox[t]{4.3cm}{Zhu, Baryakhtar, Papa, Tsuna, Kawanaka \& Eggenstein (2020) \cite{Zhu2020}} & \parbox[t]{4.0cm}{Simulated population of $\sim10^8$ isolated Galactic BHs} & \parbox[t]{1.9cm}{\centering Scalar} & \parbox[t]{5.4cm}{$2\times10^{-13}$--$2.5\times10^{-12}$~eV} \\
\hline

\parbox[t]{0.6cm}{\centering 13} & \parbox[t]{4.3cm}{Palomba, D'Antonio, Astone, Frasca, Intini, Rosa \& Leaci (2019) \cite{Palomba2019}} & \parbox[t]{4.0cm}{All-sky O2 continuous-wave search} & \parbox[t]{1.9cm}{\centering Scalar} & \parbox[t]{5.4cm}{$1.1\times10^{-13}$--$4\times10^{-13}$~eV} \\
\hline

\parbox[t]{0.6cm}{\centering 14} & \parbox[t]{4.3cm}{Arvanitaki, Baryakhtar \& Huang (2015) \cite{Arvanitaki2015}} & \parbox[t]{4.0cm}{Stellar: M33~X-7, LMC~X-1, GRO~J1655-40, Cyg~X-1, GRS~1915+105 (combined) \newline SMBH: NGC~3783, Mrk~110 \newline MCG-6-30-15, NGC~4051} & \parbox[t]{1.9cm}{\centering Scalar} & \parbox[t]{5.4cm}{Stellar: $5.2\times10^{-13}$--$2.1\times10^{-11}$~eV \newline SMBH (NGC~3783/Mrk~110): $8.4\times10^{-19}$--$1.3\times10^{-18}$~eV \newline SMBH (remaining): $1\times10^{-17}$--$6.7\times10^{-17}$~eV} \\
\hline

\parbox[t]{0.6cm}{\centering 15} & \parbox[t]{4.3cm}{Hoof, Marsh, Sisk-Reynes, Matthews \& Reynolds (2026) \cite{Hoof:2024quk}} & \parbox[t]{4.0cm}{IRAS~09149$-$6206 (SMBH) \newline M33~X-7 (stellar)} & \parbox[t]{1.9cm}{\centering Scalar} & \parbox[t]{5.4cm}{IRAS~09149$-$6206: $9.9\times10^{-20}$--$6.1\times10^{-19}$~eV \newline M33~X-7: $1.8\times10^{-13}$--$2.9\times10^{-12}$~eV} \\
\hline

\parbox[t]{0.6cm}{\centering 16} & \parbox[t]{4.3cm}{Witte \& Mummery (2025) \cite{Witte:2024drg}} & \parbox[t]{4.0cm}{Cyg X-1 (conservative) \newline GRS~1915+105 ($n\le5$)} & \parbox[t]{1.9cm}{\centering Scalar} & \parbox[t]{5.4cm}{$1.8\times10^{-13}$--$6.1\times10^{-12}$~eV} \\
\hline

\parbox[t]{0.6cm}{\centering 17} & \parbox[t]{4.3cm}{Caputo, Franciolini \& Witte (2025) \cite{caputo2025}} & \parbox[t]{4.0cm}{GW231123\_135430} & \parbox[t]{1.9cm}{\centering Scalar} & \parbox[t]{5.4cm}{$(0.6$--$5)\times10^{-13}$~eV for $f_\Phi\gtrsim10^{14}$~GeV} \\
\hline

\parbox[t]{0.6cm}{\centering 18} & \parbox[t]{4.3cm}{Collaviti, Sun, Galanis \& Baryakhtar (2025) \cite{Collaviti2024}} & \parbox[t]{4.0cm}{Cygnus X-1 } & \parbox[t]{1.9cm}{\centering Scalar} & \parbox[t]{5.4cm}{$6\times10^{-13}$--$1.4\times10^{-12}$~eV \newline(t$_{\rm age}=10^5$~yr)} \\
\hline

\parbox[t]{0.6cm}{\centering 19} & \parbox[t]{4.3cm}{Baryakhtar, Lasenby \& Teo (2017) \cite{Baryakhtar2017}} & \parbox[t]{4.0cm}{Stellar BHs \newline SMBHs} & \parbox[t]{1.9cm}{\centering Vector} & \parbox[t]{5.4cm}{$5\times10^{-14}$--$2\times10^{-11}$~eV (stellar, 2-sigma) \newline $6\times10^{-20}$--$2\times10^{-17}$~eV (SMBH, 90\% C.L.)} \\
\hline

\parbox[t]{0.6cm}{\centering 20} & \parbox[t]{4.3cm}{Ghosh \& Sachdeva (2021) \cite{Ghosh:2021zuf}} & \parbox[t]{4.0cm}{GW190517\_055101 \newline GW190426\_152155} & \parbox[t]{1.9cm}{\centering Vector} & \parbox[t]{5.4cm}{$1.7\times10^{-14}$--$7.6\times10^{-13}$~eV (GW190517\_055101) \newline $1.3\times10^{-13}$--$4.2\times10^{-12}$~eV (GW190426\_152155)} \\
\hline

\parbox[t]{0.6cm}{\centering 21} & \parbox[t]{4.3cm}{Tsukada, Brito, East \& Siemonsen (2021) \cite{Tsukada2021}} & \parbox[t]{4.0cm}{Cosmological stochastic GW background \newline (Advanced LIGO O1--O2)} & \parbox[t]{1.9cm}{\centering Vector} & \parbox[t]{5.4cm}{$0.8\times10^{-13}$--$6.0\times10^{-13}$~eV} \\
\hline

\parbox[t]{0.6cm}{\centering 22} & \parbox[t]{4.3cm}{Abac et al. (2025) \cite{LIGOScientific:2025csr}} & \parbox[t]{4.0cm}{Cygnus X-1 (O4a) \newline GW231123\_135430 remnant (O4a) \newline GW230814\_230901 remnant (O4a)} & \parbox[t]{1.9cm}{\centering Vector} & \parbox[t]{5.4cm}{$0.85$--$1.59\times10^{-13}$~eV, excluding a gap at $[1.44,1.45]\times10^{-13}$~eV (Cyg X-1, 95\% confidence) \newline $0.94$--$1.08\times10^{-13}$~eV (GW231123\_135430 remnant, 30\% confidence) \newline $2.75$--$3.28\times10^{-13}$~eV (GW230814\_230901 remnant, 30\% confidence)} \\

\end{longtable}
\end{center}
\twocolumngrid
\end{document}